\documentclass[sort&compress]{elsarticle}
\usepackage{graphicx}
\usepackage{color}
\usepackage{hyperref}
\usepackage{amsmath}
\usepackage{multirow}
\usepackage{lineno}
\newcommand{\tr}{{\rm Tr}}
\newcommand{\fdg}{{\rm diag}}

\begin{document}
\begin{frontmatter}
\title{$SU(3)_{F}$ Gauge Family Model and New Symmetry Breaking Scale From FCNC Processes}

\author[itp,sdu]{Shou-Shan Bao}
\ead{ssbao@sdu.edu.cn}
\author[itp,ucas]{Zhuo Liu}
\ead{liuzhuo@itp.ac.cn}
\author[itp,ucas]{Yue-Liang Wu}
\ead{ylwu@itp.ac.cn}
\address[itp]{%
State Key Laboratory of Theoretical Physics(SKLTP),\\
Kavli Institute for Theoretical Physics China (KITPC),
Institute of Theoretical Physics, Chinese Academy of Sciences, Beijing, 100190, PR China}
\address[ucas]{
University of Chinese Academy of Sciences (UCAS), Beijing, 100190, PR China}
\address[sdu]{School of Physics, Shandong University, Jinan, 250100, PR China}

\begin{abstract}
Based on the $SU(3)_{F}$ gauge family symmetry model which was proposed to explain the observed mass and mixing pattern of neutrinos, we investigate the symmetry breaking, the mixing pattern in quark and lepton sectors, and the contribution of the new gauge bosons to some flavour changing neutral currents (FCNC) processes at low energy. With the current data of the mass differences in the neutral pseudo-scalar $P^{0}-\bar{P}^{0}$ systems, we find that the $SU(3)_{F}$ symmetry breaking scale can be as low as 300TeV and the mass of the lightest gauge boson be about $100$TeV. Other FCNC processes, such as the lepton flavour number violation process $\mu^{-}\rightarrow e^{-}e^{+}e^{-}$ and the semi-leptonic rare decay $K\rightarrow \pi \bar{\nu} \nu$, contain contributions via the new gauge bosons exchanging. With the constrains got from $P^0-\bar{P}^0$ system, we estimate that the contribution of the new physics is around $10^{-16}$, far below the current experimental bounds.
\end{abstract}

\end{frontmatter}
\section{Introduction}

The last five decades have witnessed the great triumph of the standard model (SM). Especially the Higgs boson was finally
discovered at the Large Hadron Collider (LHC) \cite{LHC_Atlas_2012,LHC_CMS_2012}. However, there are some solid experimental evidences hinting new physics beyond SM. These evidences include neutrino oscillations~\cite{NO_SNO_PhysRevLett.87.071301,NO_K_PhysRevLett.77.1683}, dark matter (DM) \cite{DM_Markevitch:2003at,DM_Clowe:2006eq} and baryon asymmetry of the universe (BAU) \cite{BAU:1,BAU:2}. Neutrino oscillations can be explained by nonzero but tiny masses of neutrinos. And the observed nearly tri-bimaximal mixing pattern~\cite{TBM,TBM1,TBM2,TBM3,TBM4,deMedeirosVarzielas:2005qg} strongly indicates new symmetries, discrete or continuous, in the neutrino flavour sector. In general, models~\cite{deMedeirosVarzielas:2005ax,Wu:2007tq,Wu:2008ep,Wu:2012su3,YLW5,CS,MA,BHKR,Altarelli:2005yx} inhabited by these new flavour symmetries contain new heavy particles and new CP violation (CPV) phases. As a bonus, these models may provide candidates of the DM, and new CPV sources accounting for BAU. So the flavour symmetry can be a possible solution to the puzzles mentioned above.

In SM, before electroweak symmetry is spontaneously broken, quarks and leptons are all massless. Due to the universality of gauge interactions, no quantum number can distinguish
the three families. Only the Yukawa interactions can tell them
apart. Thus a simple extension to SM is to introduce a new flavour
symmetry among the three families, which is then broken
spontaneously. In this work we take the $SU(3)$ as the flavour symmetry group, denoted as $SU(3)_F$. The flavour structure of Minimal Flavour Violation in quark and lepton sectors
based on family symmetries have been
discussed in \cite{D'Ambrosio:2002ex,  Cirigliano:2005ck,Grinstein:2006cg,Buras:FCNC:2011wi,Alonso:2011jd}. Models based on other
family symmetry, such as $SO(3)_F$ symmetry, have been discussed
in \cite{Wu:2007tq,Wu:2008ep,Wetterich:1998vh,Barbieri:1999km,King:2005bj,Bao:2012nf}.

In the $SU(3)_F$ gauged family symmetry model~\cite{Wu:2012su3}, there are new interactions among the three families. The extended gauge symmetry group becomes $SU(3)_F\otimes SU(3)_c \otimes$ $SU(2)_L \otimes U(1)_Y$. As the SM Higgs field being singlet under this new family symmetry transformation, new Higgs fields are needed to break the $SU(3)_F$ symmetry. A Hermitian field $\Phi=\Phi^\dagger$ which is adjoint representation of the $SU(3)_F$ can do this job. Actually, to explain the mass
and the mixing pattern both in quark and
lepton sectors, we need two Hermitian fields $\Phi_{1,2}=\Phi_{1,2}^\dagger$. In the lepton sector, we also need right handed neutrinos $N_R$ and seesaw mechanism~\cite{Mohapatra:1979ia,Yanagida:1980xy,Schechter:1980gr} to explain the tiny neutrino masses. So there should be a complex symmetric Higgs $\Phi_\nu=\Phi_\nu^T$ to generate Majorana mass
terms for $N_R$.  The new Higgs fields transform under the $SU(3)_F$ gauge transformation as
\begin{equation}
\Phi_{1,2}\to g\Phi_{1,2}g^\dagger,\,\Phi_\nu\to g\Phi_\nu g^T, \quad g(x)\in SU(3)_F.
\end{equation}
For the representation of $SU(3)$, one has $\underline{3}\otimes \underline{3}=\underline{6}\oplus \bar{3}$ where the $\underline{6}$ representation denoted as $(2,0)$ in $p-q$ notation is symmetric while $\bar{3}$ is anti-symmetric. Here the $\Phi_\nu$ is the symmetric $\underline{6}$ representation of $SU(3)_F$.
Seesaw mechanism can also be used to explain the
mass hierarchy structures in quark and charged lepton
sectors. There could also be new heavy charged fermion fields as cousins of
$N_R$, and a new $SU(3)_F$ singlet Higgs $\phi_s$ to couple these
new heavy fields with SM fields together. We can write
down the general particle contents based on
$SU(3)_F$ gauge family symmetry with features mentioned
above, as listed in Table.\ref{tab:fermion}. For the
new gauge transformation acting in the same way on the left handed and right
handed parts of all fermions, no chiral anomaly occurs here.
\begin{table}[t]
\renewcommand{\arraystretch}{1.1}
\begin{center}
\begin{tabular}{c|c|l} 
\hline\hline
 &Fields &Representation\\
\hline
\multirow{5}*{SM fermions}&$\left(\begin{array}{c}u,c,t\\ d,s,b\end{array}\right)_L$&$\left(3_F,3_C,2_L,(1/6)_Y\right)$ \\
\cline{2-3}&$(u,c,t)_{R}$&$\left(3_F,3_C,1_L,(2/3)_Y\right)$ \\
\cline{2-3}&$(d,s,b)_{R}$&$\left(3_F,3_C,1_L,(-1/3)_Y\right)$ \\
\cline{2-3}&$\left(\begin{array}{c}e,\mu,\tau\\ \nu_e,\nu_\mu,\nu_\tau\end{array}\right)_L$&$\left(3_F,1_C,2_L,(-1/2)_Y\right)$ \\
\cline{2-3}&$(e,\mu,\tau)_{R}$&$\left(3_F,1_C,1_L,(-1)_Y\right)$ \\
\hline
SM Higgs & $H$ &$\left(1_F,1_C,2_L,(1/2)_Y\right)$\\
\hline
\multirow{4}*{New fermions}&$U$&$\left(3_F,3_C,1_L,(2/3)_Y\right)$ \\
\cline{2-3}&$D$&$\left(3_F,3_C,1_L,(-1/3)_Y\right)$ \\
\cline{2-3}&$E$&$\left(3_F,1_C,1_L,(-1)_Y\right)$ \\
\cline{2-3}&$N_R$&$\left(3_F,1_C,1_L,0_Y\right)$ \\
\hline
\multirow{3}*{New Higgs}&$\Phi_1, \Phi_2$&$\left(8_F,1_C,1_L,0_Y\right)$ \\
\cline{2-3}&$\Phi_\nu$&$\left(6_F,1_C,1_L,0_Y\right)$ \\
\cline{2-3}&$\phi_s$&$\left(1_F,1_C,1_L,0_Y\right)$ \\
\hline
\end{tabular}
\caption{The particle contents of the model with
$SU(3)_F$ gauge symmetry and their representation of gauge group $SU(3)_F\otimes
  SU(3)_C\otimes SU(2)_L\otimes U(1)_Y$. The $1_F$($1_C$, $1_L$) means that the field is singlet of $SU(3)_F$($SU(3)_C$, $SU(2)_L$) while the $0_Y$ means the hypercharge of the field is 0. The $\Phi_{1,2}$ are the Hermitian adjoint representation and the $\Phi_\nu$ is the symmetric $\underline{6}$ representation of $SU(3)_F$. $\Phi_\nu$'s VEVs produce Majorana mass terms for right handed neutrinos. $N_R$,$E$,$D$,$U$ are additional heavy fields that generate mass hierarchy structures in lepton and quark sectors.}
\end{center} \label{tab:fermion}
\end{table}

The general form of the Lagrangian is
\begin{eqnarray}
  \mathcal{L}=\mathcal{L}_{G}+\mathcal{L}_{k}+\mathcal{L}_{H}+\mathcal{L}_{Y}
  +\mathcal{L}_{n},
\end{eqnarray}
where $\mathcal{L}_{G}$ contains the kinetic and self-interaction terms of gauge bosons, including the new gauge bosons. $\mathcal{L}_{k}$ is the covariant kinetic
term of the SM fermions, and contains the new gauge
interactions among the three families's fermions mediated by the
eight new gauge bosons. And $\mathcal{L}_{H}=\mathcal{L}_{DH}-V$, with $\mathcal{L}_{DH}$
the Higgs fields' covariant kinetic terms, and $V$ the Higgs potential. $\mathcal{L}_{DH}$ gives masses to all the gauge bosons after spontaneously symmetry breaking(SSB). $V$ undergoes the SSB and gives mass terms of Higgs bosons. $\mathcal{L}_{Y}$ is the Yukawa
interactions among all the fermions and Higgs fields. It
generates masses for SM fermions and the new heavy fermions.
The new fermions' kinetic and gauge
interactions terms are collected in $\mathcal{L}_{n}$. Explicit
expressions of these terms are listed in \ref{sec:app}.

With the eight new gauge bosons, there are tree level flavour changing
neutral currents (FCNC), as well as processes that violate CP or
lepton flavour numbers. These processes are suppressed in SM. In this work we use the experimental data of these processes, to get constraints on the breaking scale of this new $SU(3)_F$ gauge symmetry.

We show the breaking pattern of the new family symmetry in Sec.\ref{sec:SSB}, and then give out the new effective Hamiltonian mediated by the new gauge bosons in Sec.\ref{sec:Effective_O}. After that the current experimental results of the neutral pseudo-scalar meson systems are used to constrain the broken scale
of this family symmetry in Sec.\ref{sec:scale}. Then we use these constraints to estimate new contributions to the semi-leptonic rare Kaon decay in Sec.\ref{sec:kaon_rare_decay} and the lepton flavour number
violating (LFNV) processes in Sec.\ref{sec:LFNV}. A short conclusion is given in Sec.\ref{sec:conclusion}.

\section{Spontaneous Breaking of the $SU(3)_F$ family symmetry}\label{sec:SSB}

Masses of the $SU(3)_F$ family gauge bosons come from their interactions with the Higgs fields $\Phi_1,\Phi_2$ and
$\Phi_{\nu}$, as described by the covariant derivative terms of $\Phi_{1,2}=\Phi_{1,2}^\dagger$ and $\Phi_\nu=\Phi_{\nu}^T$ in
$\mathcal{L}_{Higgs}$,
\begin{eqnarray}
  D_\mu \Phi_{1,2} &=& \partial_{\mu} \Phi_{1,2} -i g_F A_{F,\mu} \Phi_{1,2} +i g_F\Phi_{1,2}
  A_{F,\mu}^\dagger,\notag\\
  D_\mu \Phi_{\nu} &=& \partial_{\mu} \Phi_{\nu} -i g_F A_{F,\mu} \Phi_{\nu} -i g_F\Phi_{\nu}
  A_{F,\mu}^T.
\end{eqnarray}
The covariant kinetic terms are
\begin{eqnarray}
\label{eq:kinetic_c}
  \mathcal{L}_{DH}=\tr \Big((D_\mu \Phi_{1})(D^\mu
  \Phi_{1})^\dagger+(D_\mu \Phi_{2})(D^\mu
  \Phi_{2})^\dagger+(D_\mu\Phi_{\nu})(D^\mu\Phi_{\nu})^*\Big).
\end{eqnarray}
We use $\Phi_{1,2}$ to generate masses for quarks and charged
leptons, for only one Hermitian $\Phi$ cannot produce the observed
mixing in quark sector. And $\Phi_\nu$ generates neutrino masses
through seesaw mechanism~\cite{Schechter:1980gr}.

We assume that the vacuum expectation values (VEV) of $\Phi_\nu$ are higher than that of $\Phi_{1,2}$
and dominate the contribution to the new gauge bosons masses, since neutrinos are much lighter than the charged fermions. To show that, we use $\Phi^E$, which is a combination of $\Phi_{1,2}$,$
\Phi^{E}=[\Delta^{E}_{1}\Phi_{1}+\Delta^{E}_{2}\Phi_{2}]/\xi^e$,
to generate charge leptons masses. The corresponding Yukawa
interactions are
\begin{eqnarray}
  \label{eq:LYukawa}
\mathcal{L}_{Yukawa}&=&y_L^{e}\bar{l}H E
+y_R^e\overline{e}_R\phi_s E+\frac{1}{2}\xi^e\overline{E}\Phi^{E} E\notag\\
&&+y_L^\nu\bar{l}\tilde{H}N_R+\frac{1}{2}\xi^\nu\overline{N}_R\Phi_\nu
N_R^c+ H.c..
\end{eqnarray}

The nearly tri-bimaximal mixing pattern of neutrinos can be
explained by a residual $Z_2$ symmetry after SSB of $SU(3)_F$. 
The VEVs of the Higgs fields are assumed as the following forms~\cite{Wu:2012su3} 
\begin{eqnarray}
\label{eq:vacuum_of_higgs}
\langle H\rangle&=&v,\quad\langle\phi_s\rangle=v_s,\notag\\
 \langle\Phi^{E}\rangle&=&\Lambda_{e}\fdg(v_1,v_2,v_3)
  \Lambda_{e}^\dagger,\notag\\
\langle\Phi_\nu\rangle&=& V_{0}+\left(
\begin{array}{ccc}
 V_{1} & V_{2} & V_{2} \\
 V_{2} & V_{2} & V_{1} \\
 V_{2} & V_{1} & V_{2} \\
\end{array}
\right)=\Lambda_{\nu} \fdg(V_{1}^{\nu},V_{2}^{\nu},V_{3}^{\nu}) \Lambda_{\nu}^T,
\end{eqnarray}
where
\begin{eqnarray}
\Lambda_{\nu}=U_{TB}=
\begin{pmatrix}
 \frac{2}{\sqrt{6}} & \frac{1}{\sqrt{3}}&0\\
-\frac{1}{\sqrt{6}} & \frac{1}{\sqrt{3}}&\frac{1}{\sqrt{2}} \\
-\frac{1}{\sqrt{6}} & \frac{1}{\sqrt{3}}& -\frac{1}{\sqrt{2}}
\end{pmatrix}
\end{eqnarray}
 is the tri-bimaximal neutrino mixing matrix among three families, as a result of the residual $Z_2$ symmetry.
$V_j$ ($j=0,1,2$) is the VEV of component field of
$\Phi_\nu$, which possesses a residual $Z_2$ symmetry. After diagonalising $\langle\Phi_\nu\rangle$, we get
\begin{eqnarray}
 V_{1}^{\nu}&=&V_{0}-V_{1}+V_{2},\notag\\
  V_{2}^{\nu}&=&V_{0}+V_{1}+2V_{2},\notag\\
  V_{3}^{\nu}&=&V_{0}+V_{1}-V_{2}.
\end{eqnarray}
To get the mass eigenstates, diagonalising the mass matrices of neutrino and charged leptons as follows
\begin{equation}
U^{T}_\nu M_\nu U_\nu=diag(m_{\nu1},m_{\nu2},m_{\nu3}),\quad 
U^{\dagger}_e M_e U_e=diag(m_{e},m_{\mu},m_{\tau}) \label{eq:mixing_lepton}
\end{equation}
One has $U_\nu=U_{TB}$ due to the $Z_2$ symmetry and $U_e\sim1$ due to the approximate global $U(1)$ symmetries after spontaneous symmetry breaking~\cite{Wu:2012su3}. $U_e$ is expected to has similar hierarchy structure to Cabbibo-Kobayashi-Maskawa (CKM) mixing matrix~\cite{Kobayashi:1973fv}, which gives Pontecorvo-Maki-Nakagawa-Sakata (PMNS) matrix~\cite{Pontecorvo:1957cp,Maki:1962mu,Pontecorvo:1967fh} $U_{PMNS}=U_e^\dagger U_{TB}$ some deviation from $U_{TB}$ with non-zero $\theta_{13}$. 
One can get the mass spectrum of SM charged leptons and neutrinos are
\begin{eqnarray}
  M_e^{i}\simeq\frac{y_L^{e}y_R^{e}v v_s}{\xi^{e} v_i},\quad
   M_{\nu}^{j}\simeq\frac{(y_L^{\nu}v)^2}{\xi^{\nu} V_{j}^{\nu}},
\end{eqnarray}
where the index $i=1,2,3$ stands for charged leptons mass eigenstates
$e,\mu,\tau$. And $j=1,2,3$ stand for neutrinos mass eigenstates
$\nu_1,\nu_2,\nu_3$. The observed neutrinos' mass hierarchy suggests
$V_{0}\gg V_{1},V_{2}$. Since $m_e\ll m_\mu<m_\tau$, there should be $v_1\gg v_2>v_3$.

Taking all the Yukawa couplings to be nature and of order $1$, we get
their masses are
\begin{eqnarray}
  M_{e}^{i}\sim \frac{v v_s}{v_i}, \quad
    M_{\nu}^{j}\sim \frac{v^2}{V_{j}^{\nu}}\sim\frac{v^2}{V_{0}}.
\end{eqnarray}
Assuming $M_{\nu}^{j}\sim 0.1$eV and using $m_e\sim 0.5$MeV we can
get $V_0\sim10^{14}$GeV, $v_1\sim 10^{5}v_s$. The Yukawa
couplings can be tuned to reduce all the scales. With $\xi^e,\xi^{\nu}\sim 1$,
tuning $y_{L}^e,y_{R}^e\sim 10^{-2}$ and
$y_{L}^\nu,y_{R}^\nu\sim 10^{-4}$, we get $v_1\sim 10v_s$,
$V_{0}\sim 10^3$TeV. With the assumption that
$v_s\sim $TeV, there is $|V_{0}|\gg  v_1$. So we can
safely neglect contribution from $\langle\Phi_{1,2}\rangle$ in
Eq.(\ref{eq:kinetic_c}) and only consider that from $\langle\Phi_{\nu}\rangle$. There is another benefit for this interval of $v_s$'s value. The Higgs field $\phi_s$ can mixing
with the SM Higgs field and be a cold dark matter candidate. Neglecting $\langle\Phi_1\rangle,\langle\Phi_2\rangle$ in
Eq.(\ref{eq:kinetic_c}), we get
 \begin{eqnarray}
  \mathcal{L}\supset g_{F}^2 \tr\left(A_{F}^{\mu}\Phi_{\nu}A_{F,\mu}^{*}\Phi_{\nu}^{*}
  +A_{F}^{\mu}\Phi_{\nu}\Phi_{\nu}^{*}A_{F,\mu}^{\dagger}
  +\Phi_{\nu}A_{F,\mu}^{T}A_{F}^{\mu *}\Phi_{\nu}^{*}
  +\Phi_{\nu}A_{F,\mu}^{T}\Phi_{\nu}^{*}A_{F}^{\mu\dagger}\right).\notag
\end{eqnarray}

In the following parts of this paper, we denote $A^a_{F,\mu}, A_{F,\mu}^{a}T^{a}$ as
$F^a_\mu,F_{\mu}$ for short. They can be parameterised by the Gell-Mann
matrices with $T^a=\lambda^a/2$,
\begin{eqnarray}
  F_\mu=F_{\mu}^{a}\frac{\lambda^a}{2}=\left(
\begin{array}{ccc}
 \frac{1}{2} \left(F_3+\frac{F_8}{\sqrt{3}}\right) & \frac{1}{2} \left(F_1-i
   F_2\right) & \frac{1}{2} \left(F_4-i F_5\right) \\
 \frac{1}{2} \left(F_1+i F_2\right) & \frac{1}{2}
   \left(\frac{F_8}{\sqrt{3}}-F_3\right) & \frac{1}{2} \left(F_6-i F_7\right) \\
 \frac{1}{2} \left(F_4+i F_5\right) & \frac{1}{2} \left(F_6+i F_7\right) &
   -\frac{F_8}{\sqrt{3}}
\end{array}
\right)_{\mu}.
\end{eqnarray}
The gauge family symmetry breaks down to residual $Z_2$ symmetry with non-zero $V_{0,1,2}$. If $V_{0}\neq0$ and  $V_{1}=V_{2}=0$, the $SU(3)_F$ symmetry is broken down to $SO(3)_F$
symmetry. Then there are 5 gauge family fields, $F_1,F_3,F_4,F_6$ and $F_8$, gaining
degenerate masses $m=2g_FV_{0}$. The other 3 fields $F_2,F_5,F_7$, which
corresponding to the unbroken $SO(3)_F$ symmetry, remain massless.
The $SO(3)_F$ is besides broken with non-zero $V_{1,2}$ and a $Z_2$ symmetry is left.
The masses of $F_2,F_5,F_7$ are smaller comparing with the
other five since $V_{1,2}<V_0$. We denote that
\begin{eqnarray}
  \label{eq:alpha_beta}
   \frac{V_{1}}{V_{0}}\equiv {r_1} ,\quad
  \frac{V_{2}-V_{1}}{V_{0}}\equiv{r_2},\label{eq:v0}
\end{eqnarray}
and assume ${r_1}$ and ${r_2}$ are of same order of the Wolfenstein parameter $\lambda\sim0.22$. A detailed analysis of neutrinos mass spectrum~\cite{Wu:2012su3} shows ${r_1}\sim
\lambda,{r_2}\sim \mp 2\lambda$ can be used to explain the normal
and inverted mass hierarchy spectrum of left handed neutrinos. We
can use $V_{0},V_{1},V_{2}$, or equally $V_{0},{r_1},{r_2}$ to get
the mass spectrum of the new family gauge bosons. With the
abbreviations
\begin{eqnarray}
 \mathcal{F}_{5}&=&\left(F_1,F_3,F_4,F_6,F_8\right)^T,\quad
 \mathcal{F}_{3}=\left(F_2,F_5,F_7\right)^T,
\end{eqnarray}
the mass terms can be expressed as
\begin{eqnarray}
  \label{eq:massterm}
  \mathcal{L}_{mass}= g^{2}_{F}V_{0}^2\mathcal{F}_5^T \left(\mathcal{M}_{5\times
  5}^{2}+\delta\mathcal{M}_{5\times
  5}^{2}\right)\mathcal{F}_5
  +  g^{2}_{F}V_{0}^2\mathcal{F}_3^T \left(\mathcal{M}_{3\times
  3}^{2}\right)\mathcal{F}_3,
\end{eqnarray}
where the matrices are
\begin{eqnarray}
    \mathcal{M}_{5\times
  5}^{2}&=& \left(
\begin{array}{ccccc}
 r_0 & 0 & 2 {r_1}  & 2 {r_1} +2 {r_2}  & \frac{4 {r_1}
   }{\sqrt{3}}+\frac{4 {r_2} }{\sqrt{3}} \\
 0 & r_0 & 2 {r_1} +2 {r_2}  & -2 {r_1}  & -\frac{2 {r_2}
   }{\sqrt{3}} \\
 2 {r_1}  & 2 {r_1} +2 {r_2}  & r_0 & 2 {r_1} +2 {r_2}  &
   -\frac{2 {r_1} }{\sqrt{3}}-\frac{2 {r_2} }{\sqrt{3}} \\
 2 {r_1} +2 {r_2}  & -2 {r_1}  & 2 {r_1} +2 {r_2}  & {r_0} +2{r_2} 
   & -\frac{2 {r_1} }{\sqrt{3}} \\
 \frac{4 {r_1} }{\sqrt{3}}+\frac{4 {r_2} }{\sqrt{3}} & -\frac{2 {r_2}
   }{\sqrt{3}} & -\frac{2 {r_1} }{\sqrt{3}}-\frac{2 {r_2} }{\sqrt{3}} &
   -\frac{2 {r_1} }{\sqrt{3}} &  {r_0} +\frac{4 {r_2} }{3} \\
\end{array}
\right)\notag\\
\label{eq:massMatrix1}
\end{eqnarray}
with $r_0\equiv 2+4r_1+2r_2$, and
\begin{eqnarray}
 \label{eq:massMatrix11}
&&\delta\mathcal{M}_{5\times
  5}^{2}=\delta\mathcal{M}_{5\times
  5}^{2}({r_1}^2,{r_1}{r_2},{r_2}^2) \sim \mathcal{O}(\lambda^2),
\end{eqnarray}
\begin{flalign}
  \label{eq:massMatrix2}
&\mathcal{M}_{3\times
  3}^{2}=\notag\\
  &\left(
\begin{array}{ccc}
 3 {r_1} ^2+5 {r_2}  {r_1} +3 {r_2} ^2 & \frac{3 {r_1} ^2+8 {r_2}  {r_1} +3 {r_2} ^2}{2}  & \frac{{r_2} ^2-3 {r_1} ^2-2 {r_2}{r_1}}{2} \\
 \frac{3 {r_1} ^2+8 {r_2}  {r_1} +3 {r_2} ^2}{2}  & 3 {r_1}
   ^2+5 {r_2}  {r_1} +3 {r_2} ^2 &\frac{3{r_1} ^2+2{r_2}  {r_1} -{r_2} ^2}{2}  \\
 \frac{{r_2} ^2-3 {r_1} ^2-2 {r_2}  {r_1}}{2}  & \frac{3
   {r_1} ^2+2{r_2}  {r_1} -{r_2} ^2}{2} & 3 {r_1} ^2+2 {r_2}
   {r_1} +{r_2} ^2 \\
\end{array}
\right).
\end{flalign}
The matrix elements of $\delta\mathcal{M}_{5\times
  5}^{2}$ and $\mathcal{M}_{3\times3}^{2}$ are the of same order.
  The $\mathcal{M}_{3\times 3}^{2}$ and $\mathcal{M}_{5\times 5}^{2}$ can be diagnosed,
\begin{eqnarray}
  \label{eq:massEig2}
 \hat{\mathcal{M}^{2}}_3=u_{TB}^T \mathcal{M}_{3\times
  3}^{2} u_{TB},\quad
\hat{\mathcal{M}^{2}}_5=U_{5}^T \mathcal{M}_{5\times 5}^{2} U_{5}
  \end{eqnarray}
where $u_{TB}$ and $U_5$ are the mixing matrices
\begin{eqnarray}
u_{TB}=\left(
\begin{array}{ccc}
 \frac{1}{\sqrt{3}} & \frac{1}{\sqrt{2}} &
   -\frac{1}{\sqrt{6}} \\
 -\frac{1}{\sqrt{3}} & \frac{1}{\sqrt{2}} &
   \frac{1}{\sqrt{6}} \\
 \frac{1}{\sqrt{3}} & 0 & \sqrt{\frac{2}{3}} \\
\end{array}
\right).
\end{eqnarray}
The analytical form of  mixing matrix $U_5$ is too complex to list
here. If we take the assumption ${r_1}\sim \lambda$, and
${r_2}\sim-2\lambda$(${r_2}\sim2\lambda$) for normal hierarchy (inverted hierarchy), the
numerical results are
\begin{eqnarray}
  U^{NH(IH)}_5=\left(
\begin{array}{ccccc}
 -\frac{1}{\sqrt{3}} & \frac{1}{3 \sqrt{2}} & \frac{1}{\sqrt{6}} & -0.613(0.486) &
   0.261(-0.456) \\
 -\frac{1}{2 \sqrt{3}} & \frac{1}{\sqrt{2}} & -\frac{1}{\sqrt{6}} & 0.400(0.114) &
   0.301(-0.487) \\
 \frac{1}{\sqrt{3}} & \frac{1}{3 \sqrt{2}} & -\frac{1}{\sqrt{6}} & -0.613(0.486) &
   0.261(-0.456) \\
 0 & -\frac{\sqrt{2}}{3} & 0 & 0.186(0.714) & 0.862(0.518) \\
 \frac{1}{2} & \frac{1}{\sqrt{6}} & \frac{1}{\sqrt{2}} & 0.230(0.066) & 0.174(0.281) \\
\end{array}
\right).
\end{eqnarray}
It's notable that although the mass eigenvalues depend on
${r_1},{r_2}$, the mixing matrix $u_{TB}$ do not, which is
guaranteed by the residual $Z_2$ symmetry. With
$\delta\mathcal{M}_{5\times 5}^{2}$ treated as perturbation, we
get the mass eigenstates of the family gauge bosons
\begin{eqnarray}
  \mathcal{Z}_{5}&=&\fdg(Z_1,Z_2,Z_3,Z_4,Z_5)=U_5^{T}
  \mathcal{F}_5,\notag\\
  \mathcal{Z}_{3}&=&\fdg(Z_6,Z_7,Z_8)=u_{TB}^{T} \mathcal{F}_3.
\end{eqnarray}
The masses of the five heavy gauge bosons are
\begin{eqnarray}
 M_1&=&2g_{F}V_{0},\quad
 M_2=2g_{F}(V_{0}^{2}+2V_{1}V_{0}+V_{2}V_{0})^{1/2},\notag\\
 M_3&=&2g_{F}(V_{0}^{2}+3V_{2}V_{0})^{1/2},\notag\\
 M_4&=&\frac{2\sqrt{3}}{3}g_{F}V_{0}^{1/2}\left(2V_{0}+2V_{1}+4V_{2}+2\sqrt{4V_{1}^{2}-2V_{1}V_{2}+7V_{2}^2}\right)^{1/2},\notag\\
 M_5&=&\frac{2\sqrt{3}}{3}g_{F}V_{0}^{1/2}\left(2V_{0}+2V_{1}+4V_{2}-2\sqrt{4V_{1}^{2}-2V_{1}V_{2}+7V_{2}^2}\right)^{1/2}.
\end{eqnarray}
And the masses of the three light gauge bosons, which are related
to the $SO(3)_F$ symmetry, are
\begin{eqnarray}
  M_6=2g_{F}|V_{2}-V_{1}|,\quad M_7=3g_{F}|V_{2}|,\quad
  M_8=g_{F}|2V_{1}+V_{2}|.
\end{eqnarray}

\section{Low Energy Effective Hamiltonian}\label{sec:Effective_O}

 In general the family eigenstates of the fermions are different from weak eigenstates.
 After the SSB of $SU(3)_F$ family symmetry, the  interactions between the new family gauge bosons and SM fermions
 are
\begin{eqnarray}
\label{eq:int}
   \mathcal{L}_{int}\supset&&
   g_F\left[\overline{u}_{L}\gamma^\mu (U_{L}^{u\dagger}
  F_{\mu} U_{L}^{u})u_{L}+\overline{u}_{R}\gamma^\mu (U_{R}^{u\dagger}
  F_{\mu} U_{R}^{u})u_{R}\right]\notag\\
&+& g_F\left[\overline{d}_{L}\gamma^\mu (U_{L}^{d\dagger}
  F_{\mu} U_{L}^{d})d_{L}+\overline{d}_{R}\gamma^\mu( U_{R}^{d\dagger}
  F_{\mu} U_{R}^{d})d_{R}\right]\notag\\
  &+& g_F\left[\overline{e}_{L}\gamma^\mu( U_{L}^{e\dagger}
  F_{\mu} U_{L}^{e})e_{L}+\overline{e}_{R}\gamma^\mu (U_{R}^{e\dagger}
  F_{\mu} U_{R}^{e})e_{R}\right]\notag\\
  &+& g_F \overline{\nu}_{L}\gamma^\mu (U_{L}^{\nu\dagger}
  F_{\mu} U_{L}^{\nu})\nu_{L},
\end{eqnarray}
where all the fermion triplets are weak eigenstates, and the
corresponding mixing matrices are the clashes between weak
eigenstates and family eigenstates.
All the mass matrices of quarks and charged leptons are gained
through SM Higgs $H$ and $\Phi_{1,2}$, which are hermitian. Assuming
 all the Yukawa couplings to be real, as the situation in models with
spontaneous CP violation, we get hermitian mass matrices, and
the SSB of the new gauge symmetry and seesaw mechanism give out
\begin{eqnarray}
&& U_{L}^u=U_{R}^u=U^u,\quad U_{L}^d=U_{R}^d=U^d, \quad\notag\\
&& U_{L}^e=U_{R}^e=U_e,\quad U^{\nu}_{L}=U_{\nu}=U_{TB},
\end{eqnarray}
where $U_{e},U_{\nu}$ are the mixing matrices in Eq.(\ref{eq:mixing_lepton}) and $U^u, U^d$ are similar to $U_e$.
The mixing matrices satisfy that
\begin{eqnarray}
  \label{eq:CKMs}
    U_{CKM}=U^{u\dagger}U^{d},\quad U_{PMNS}=U^{\dagger}_{e}U_{TB},
\end{eqnarray}
Experimental measurement shows that the deviation between $U_{MNSP}$ and $U_{TB}$ is small. So we can take $U_{e}\sim 1$ as the leading-order approximation. Hence the charged lepton mass eigenstates are coincident with the family eigenstates.

All the mixing matrices are physical and can be measured via the interactions among SM fermions and $SU(3)_F$ gauge bosons. It's quite different from that in SM, where $U^u,U^d$ and $U_{e}, U_{TB}$ are not all observable, only their clashes $U_{CKM}$ and $U_{PMNS}$ hold physical meanings.

We can also assume that $U^u,U^d$ and $U_{e}$ have the same
hierarchy structures as $U_{CKM}$ and can be parameterised via
Wolfenstein method~\cite{Wolfenstein:1983yz}
\begin{eqnarray}
U_{CKM}
&\sim&\left(
\begin{array}{ccc}
 1-\frac{\lambda ^2}{2} & \lambda  & A \lambda ^3\rho e^{-i\delta} \\
 -\lambda  & 1 & A \lambda ^2 \\
 A \lambda ^3 ( 1-\rho e^{i\delta}) & -A \lambda ^2 & 1 \\
\end{array}
\right)+\mathcal{O}(\lambda^4).
\end{eqnarray}
For $U_{e}$, we replace $A,\lambda,\rho,\delta$ by
$A_e,\lambda_e,\rho_e,\delta_e$. A detailed analysis of the allowed values of
these parameters and the CP violation phases can be find in~\cite{Liu:2014}. For the mixing matrix in up(down) quark sectors, we have mixing matrix $U^u$($U^d$) with the parameters
$A,\lambda,\rho,\delta$ replaced by $A_u,\lambda_u,\rho_u,\delta_u$ $(A_d,\lambda_d,\rho_d,\delta_d)$.
Eq.(\ref{eq:CKMs}) gives out the relations of the Wolfenstein
parameters in $U_{CKM}$, $U^u$ and $U^d$ as follows,
\begin{eqnarray}
  \lambda&\sim&
  (\lambda_d-\lambda_u)(1-\frac{\lambda_d\lambda_u}{2}),\notag\\
    A\lambda^2&\sim&
  A_d\lambda_{d}^2-A_u\lambda_{u}^2,\notag\\
  e^{-i \delta}&\sim&
  \frac{A_d \lambda _d^3 \rho_d e^{-i\delta_d}-A_d \lambda _d^2 \lambda _u
  +A_u \lambda _u^3\rho_u(1-e^{-i\delta_u})}
   {A_d \lambda _d^3\rho_d-A_d \lambda _d^2 \lambda _u-A_u \lambda _d \lambda
   _u^2+A_u \lambda _u^3\rho_u}.
\end{eqnarray}
It's known that the SM Dirac CP phase $\delta$ is
not enough to generate the observed BAU~\cite{Shaposhnikov:1987pf,Gavela:1994ds,Gavela:1994dt}. And the new Dirac CP phases $\delta_{e},\delta_{u},\delta_{d}$ may help to solve the baryogenesis problem.

The low energy effective Hamiltonian mediated by these new family gauge bosons
can be written down easily,
\begin{eqnarray}
 \mathcal{H}_{eff}=\frac{1}{S}\sum_{M,N}\sum_{a,b,c}\mathcal{C}(\mu)\xi_{ij,a}^{M}\xi_{kl,c}^{N}
  \frac{g^2_F\mathcal{V}_{ab}\mathcal{V}_{cb}}{M_{b}^2}\mathcal{O}_{ij}^{M}
  \otimes\mathcal{O}_{kl}^{N} + h.c.,
\end{eqnarray}
where $i,j,k,l=1,\ldots,3$ are the indices of fundamental
representation of $SU(3)_F$, while $a,b,c=1,\ldots,8$ are the indices
of adjoint representation of $SU(3)_F$ and $M_b$ is the mass of the corresponding gauge boson. $M,N=\{u,d,e,\nu\}$ stand for the fermion's species. $S$ is the symmetric factor, $S=2$ for
$\mathcal{O}_{ij}^{M}$ and $\mathcal{O}_{kl}^{N}$ being the same,
and $S=1$ for other situations. $\mathcal{C}(\mu)$ are the Wilson
coefficients. One can find the QCD corrections at one loop level are of order
$\sim10\%$~\cite{Buras:2012fs}, at the same order of corrections when we neglect
the contributions of $\Phi_{1,2}$ to the new gauge bosons masses.
We do not consider the corrections of the Wilson coefficients in this work. The current operators are
\begin{eqnarray}
 \mathcal{O}_{ij}^{N}=\overline{N}_{i}\gamma_{\mu}{N}_{j}.
\end{eqnarray}
And the coefficients are
\begin{eqnarray}
  \xi_{ij,a}^{N}=[U^{N\dagger}T^a U^{N}]_{ij}.
\end{eqnarray}
Mixing matrix among $SU(3)_F$ gauge bosons is a block diagonal
matrix made up by $U_{5\times5}$ and $u_{TB}$,
\begin{eqnarray}
  \mathcal{V}_{ab}=[U_{5\times5}\oplus u_{TB}]_{ab}.
\end{eqnarray}

Quite a lot of effective operators occur. To suppress these new operators'
contribution, we expect that the new energy scale $V_{0},V_{1},V_{2}\gg
v\sim 173 GeV$. There are also some FCNC operators
which are absent in
SM at tree level. Such operators can contribute to the processes including the
$P^0$-$\bar{P}^0$ mixing in neutral meson systems, as well as some
LFNV processes and some
CPV processes. These processes appear in SM at loop
level through penguin diagrams and box diagrams, and are suppressed
comparing with the tree level processes. The new gauge
bosons can contribute to these processes at tree level directly. So we may find
hints of these new gauge bosons in these interesting processes.
In the following parts we will find the constraints given by these
processes respectively.

\section{Mass difference of $P^0-\bar{P}^0$}
\label{sec:scale}

In neutral meson systems, $P^0$ can mix with $\bar P ^0$, where
$P^0$ refers to either $K^0$, $D^0$, $B_d^0$ or $B_s^0$. Such mixing violates CP symmetry and has been studied widely~\cite{Gaillard:1974hs,Pich:1985ab,Datta:1984jx,Hagelin:1981zk,Franco:1981ea,Buras:1984pq}. We take
$K^0$-$\bar K^0$ as an example. In SM, $K^0$ and $\bar K ^0$ are
mixed by $\Delta S=2$ interactions through box diagrams~\cite{Niyogi:1979wm}. The
measured tiny mass difference between $K_L^0$ and $
K_S^0$~\cite{PDG_2012} puts stringent constraints on tree level FCNC beyond SM.
 The $SU(3)_F$ family gauge bosons and their
mixing can contribute to this process at tree level.
So the measured mass difference can give hint of the new gauge bosons' masses.

All the eight new gauge bosons can contribute to this mass
difference. Noticed that $Z_{6},Z_7,Z_8$ are lighter than the
other 5 gauge bosons, we may ignore the heavy ones and focus on
these lighter ones. This approximation makes
$\mathcal{V}\sim u_{TB}$. The form of $U_5$ is not concerned here.

The mass difference between $K^0$ and $\bar K ^0$ can be
calculated using methods in~\cite{Chau:1982da,Grossman:1997xn,Bao:2012nf}. The
Hamiltonian can be written as $\mathcal{H}=\mathcal{H}_0+\mathcal{H}_2$, with $\mathcal{H}_0$
refers to the strong and electromagnetic interaction parts, which
conserves the strange number. And $\mathcal{H}_2$ is the weak interaction
term and induces $\Delta S=2$ processes. The real parts of
eigenvalues of $\mathcal{H}$ are denoted as $m_L,m_S$.
Their mass difference is
\begin{eqnarray}
\label{eq:msdiff}
  \Delta m=m_L-m_S=Re\left[\langle K^0|H_2|\bar{K}^0\rangle+
    \langle\bar{K}^0|H_2|K^0\rangle\right]/(2m_K)
\end{eqnarray}

The new low energy effective Hamiltonian responsible for
$K-\bar{K}$ mixing is
\begin{eqnarray}
\label{eq:eff_H_K}
  \mathcal{H}_K^{New}&=\mathcal{C}_{K}(\bar{s}\gamma_{\mu}d)\otimes
  (\bar{s}\gamma^{\mu}d)+H.c. .
\end{eqnarray}
Here we treat $\lambda_d$ as a small parameter and get the
coefficient in Eq.(\ref{eq:eff_H_K}) to the order of
$\lambda_d^2$. At higher order the heavy family gauge bosons'
effects should be take into consideration. The coefficient
$\mathcal{C}_{K}$ is
\begin{eqnarray}
  \mathcal{C}_{K}=\frac{1}{16}[F_{K}(V_{1},V_{2})+G_{K}(V_{1},V_{2})A_d\lambda_{d}^2]
  +\mathcal{O}(\lambda_d^3),
\end{eqnarray}
where
\begin{eqnarray}
  F_{K}(V_{1},V_{2})&=&\frac{1}{6 \left(V_2-V_1\right){}^2}+\frac{1}{3 \left(2
   V_1+V_2\right){}^2}+\frac{1}{9 V_2^2}.\notag\\
 G_{K}(V_{1},V_{2})&=&\frac{1}{3 \left(V_2-V_1\right){}^2}+\frac{2}{3 \left(2
   V_1+V_2\right){}^2}-\frac{2}{9 V_2^2}.
\end{eqnarray}
The contribution of $G_{K}(V_{1},V_{2})$ are at order of $\lambda_{d}^2$.
If we assume $\lambda$ and $\lambda_d$ are of the same order,
then the contribution of $G_{K}(V_{1},V_{2})$ can be omitted as the contributions of the heavy gauge bosons. This approximation is equivalent to setting the mixing matrix $U^d\sim 1$.

To get the matrix element $\langle\bar{K}^0|\mathcal{O}|K^0\rangle$, we use the
vacuum insertion approximation (VIA). The result is
\begin{eqnarray}
\langle \bar{K}^0|(\bar{s}\gamma^{\mu}d)\otimes(\bar{s}\gamma^{\mu}d)|K^0\rangle=\frac{2}{3}N_{1}+\frac{1}{3}N_{2},
\end{eqnarray}
where~\cite{McWilliams:1980cz}
\begin{eqnarray}
    N_{1}&\equiv& \langle\bar{K}^0|
  \bar{s}\gamma^5 d|0\rangle
  \langle 0|
  \bar{s}\gamma^{5}d|K^0\rangle,\notag\\
      N_{2}&\equiv& \langle\bar{K}^0|
  \bar{s}\gamma_{\mu}\gamma_{5}d|0\rangle
  \langle 0|
  \bar{s}\gamma^{\mu}\gamma_{5}d|K^0\rangle.
  \end{eqnarray}
With the definition of Kaon decay constant $f_{K}$,
 \begin{eqnarray}
 \langle 0|\bar{s}\gamma^{\mu}\gamma_5
 d|K^0(p)\rangle
 =if_{K}p^{\mu},
 \end{eqnarray}
we get
\begin{eqnarray}
  N_1=\frac{f_{K}^{2}m_{K}^4}{(m_s+m_d)^2},\quad
  N_2=f_{K}^{2}m_{K}^{2}.
\end{eqnarray}
To the lowest order of $\lambda_d$,
\begin{eqnarray}
  \langle\bar{K}^0|{\mathcal{H}_2^{New}}|{K^0}\rangle&=&
\frac{F_{K}(V_{1},V_{2})}{16 }
  \frac{f_K^{2}M_K^{2}}{6M_K}\big[1+2\frac{M_{K}^2}{(m_s+m_d)^2}\big]\notag\\
  &=&
  \frac{F_{K}(V_{1},V_{2})f_K^{2}M_K}{96 }\big[1+2R(\mu)\big].
\end{eqnarray}
The hadronic matrix uncertainties will modify the relation
above~\cite{Buchalla:1995vs,Buras:2012fs}. From Eq.(\ref{eq:msdiff}), the new family interaction contributes to the mass difference via a new term in addition to that in SM as
\begin{eqnarray}
  \Delta m^{New}=\frac{F_{K}(V_{1},V_{2})f_K^{2}M_K}{48}\big[1+2R(\mu)\big].
\end{eqnarray}
If the new contribution saturate the mass difference, then
\begin{eqnarray}\label{vev}
 \frac{1}{ F_{K}(V_{1},V_{2})}&\geq&\frac{f_K^{2}M_K}{48\Delta m^{New}}\big[1+2R(\mu)\big]\sim
  \frac{f_K^{2}M_K}{48\Delta m_{K}}.
\end{eqnarray}

\begin{figure}
\scalebox{0.4}{\includegraphics{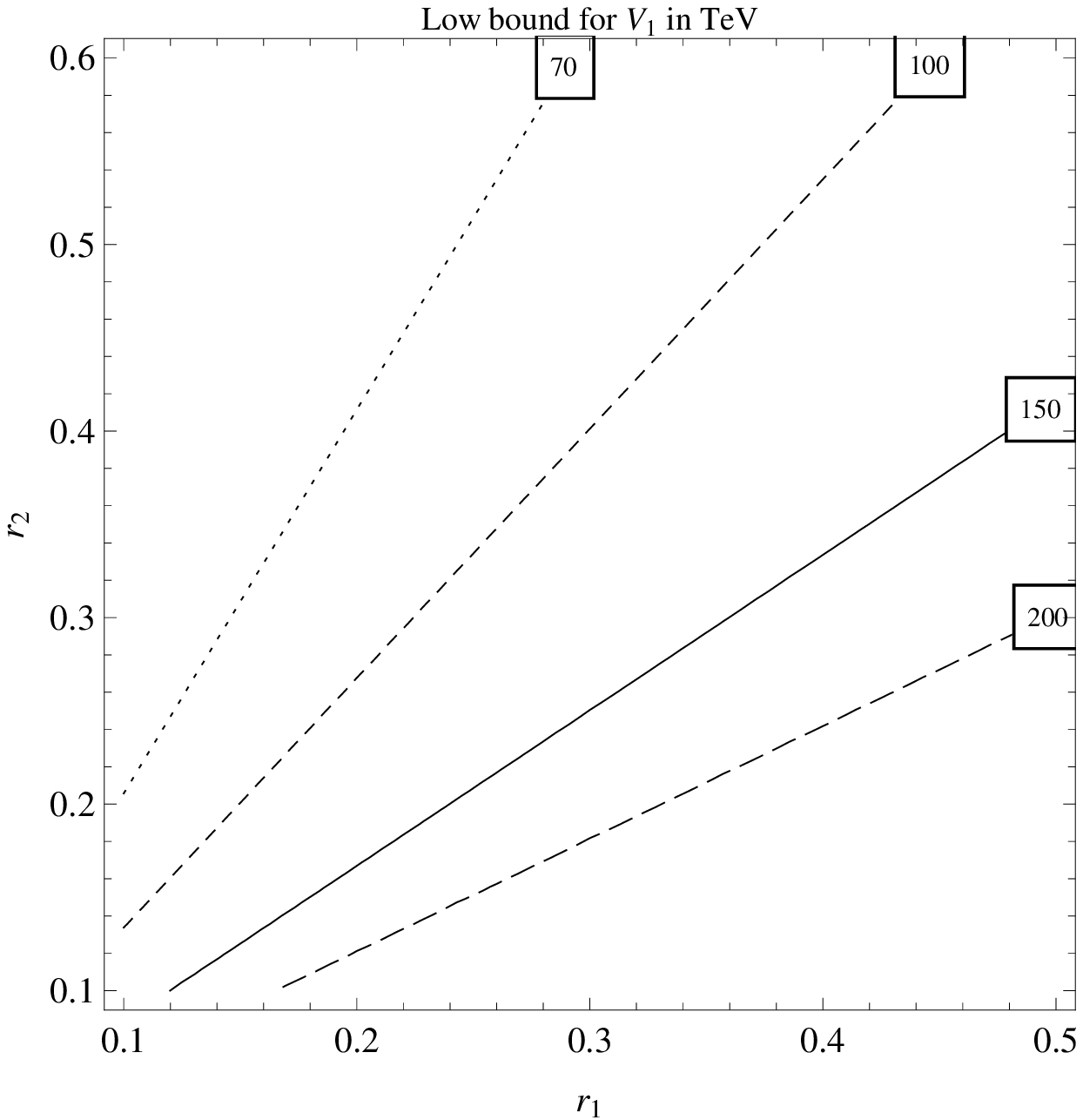}}
\scalebox{0.4}{\includegraphics{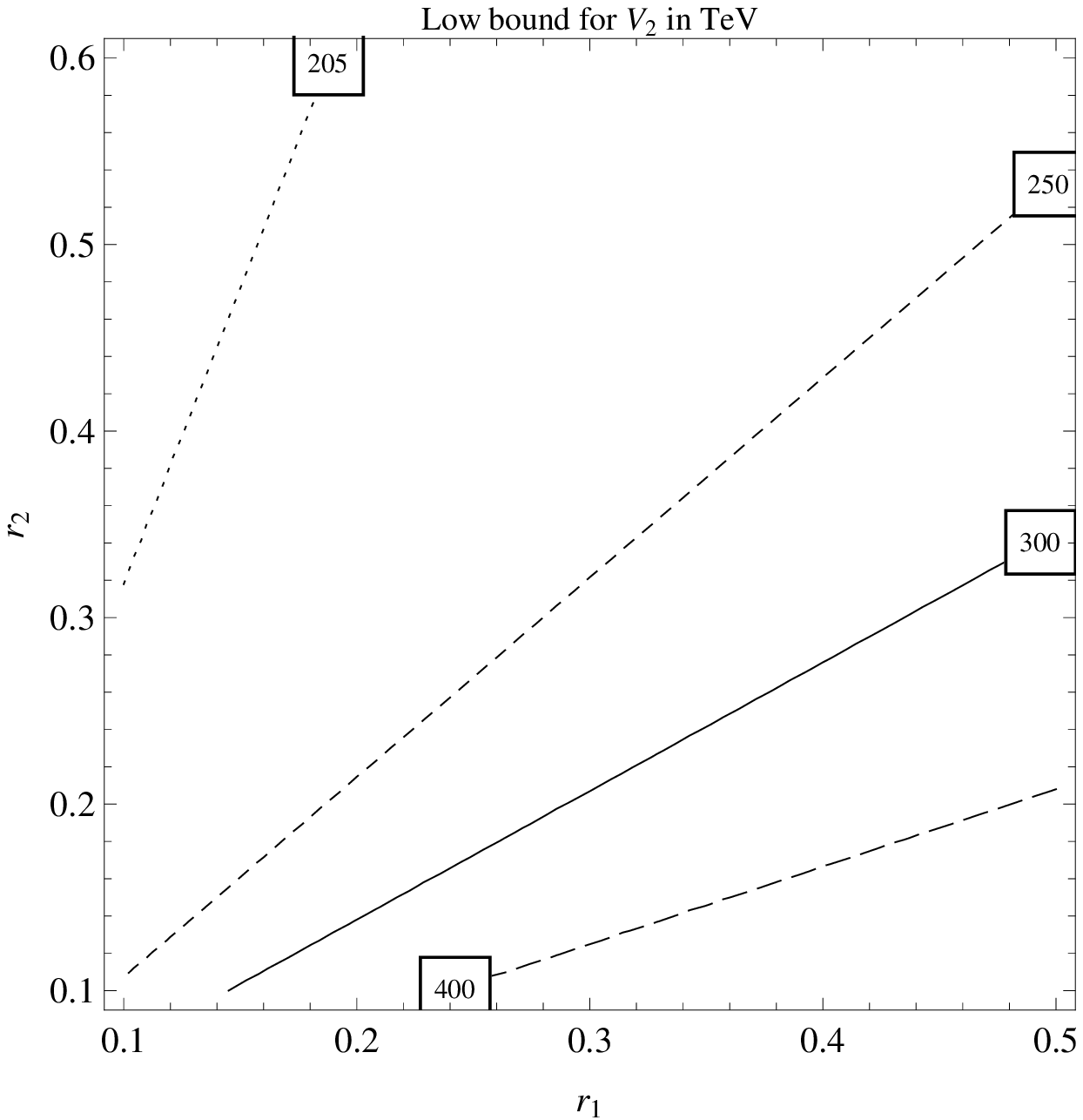}}
\scalebox{0.4}{\includegraphics{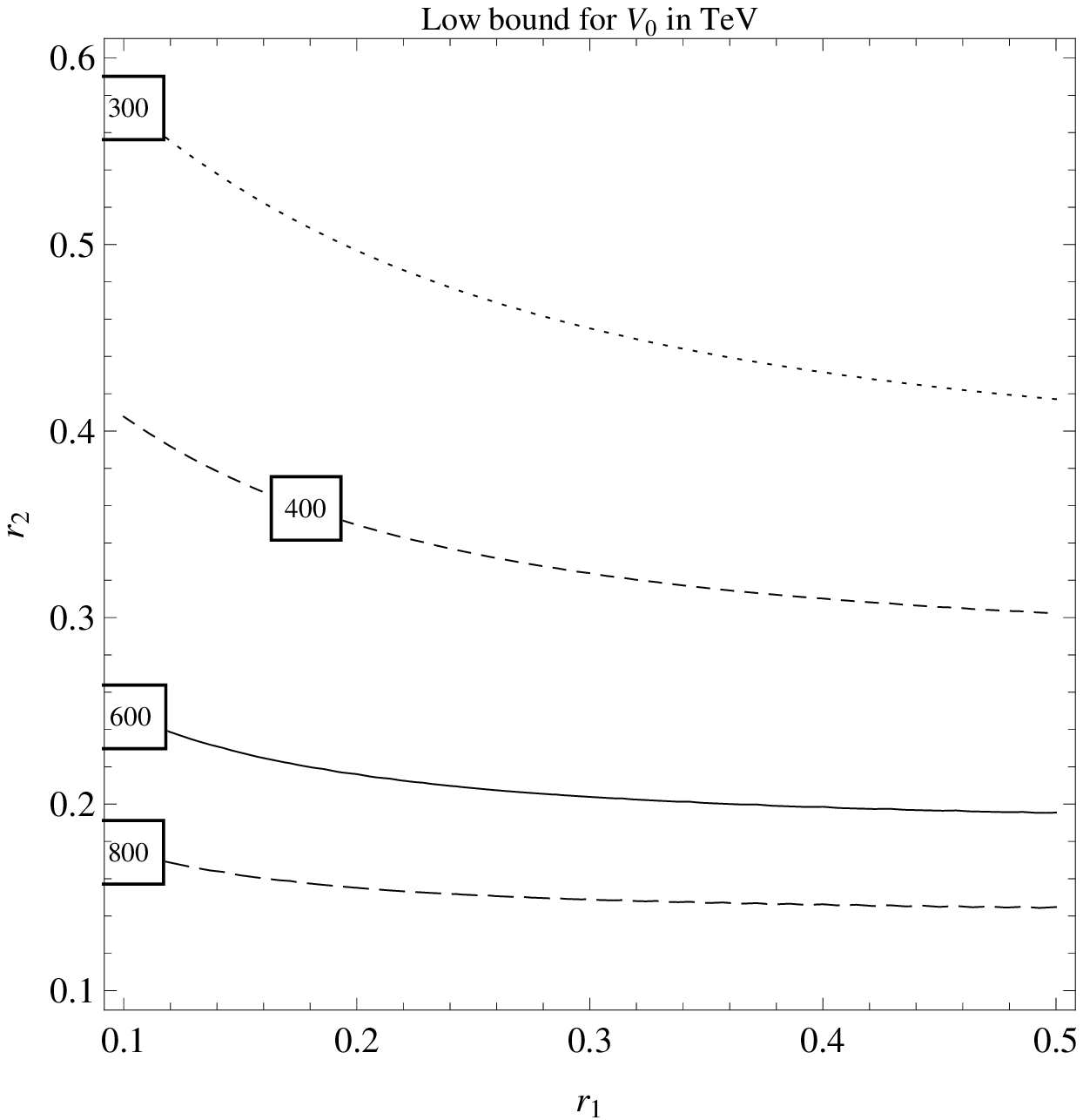}} \caption{The lower bounds of $SU(3)_F$ breaking scale $V_{1},V_{2}$ and $V_{0}$ in $TeV$
given by neutral Kaon system with different ${r_1}$ and
${r_2}$.}\label{fig:LFV}
\end{figure}
With Eq.(\ref{eq:alpha_beta}), it's easy to get
\begin{eqnarray}
  V_{1}^2&\geq & \frac{f_K^{2}M_K}{864\Delta
  m_{K}}\left[
\frac{3 {r_1} ^2}{{r_2} ^2}+\frac{2 {r_1} ^2}{({r_1} +{r_2}
)^2}+\frac{6 {r_1}
   ^2}{(3 {r_1} +{r_2} )^2}\right].
\end{eqnarray}

Using the experimental data~\cite{PDG_2012,fK_Rosner:2010ak} listed in Table.\ref{tab:res},
and taking the assumption that ${r_1}\sim \lambda$ and ${r_2}\sim
2\lambda$, we can get the bounds of the symmetry broken scales which are about
\begin{eqnarray}
  V_{1}\ge 69.8TeV, \quad V_{2}\ge 209 TeV, \quad V_{0}\ge
  317TeV.
\end{eqnarray}
The lower bounds of $V_0,V_1$ and $V_2$ as functions of ${r_1},{r_2}$ are shown in Fig.\ref{fig:LFV}. A similar analysis can be
carried out in $D-\bar{D}$, $B-\bar{B}$ and $B_s-\bar{B_s}$
systems. The effective Hamiltonian terms at the lowest order are
\begin{eqnarray}
\label{eq:eff_DB}
     \mathcal{H}_D^{New}&=&\mathcal{C}_{D}(\bar{u}\gamma_{\mu}c)\otimes
  (\bar{u}\gamma^{\mu}c),\notag\\
    \mathcal{H}_{B_d}^{New}&=&\mathcal{C}_{B_d}(\bar{b}\gamma_{\mu}d)\otimes
  (\bar{b}\gamma^{\mu}d),\notag\\
    \mathcal{H}_{B_s}^{New}&=&\mathcal{C}_{B_s}(\bar{b}\gamma_{\mu}s)\otimes
  (\bar{b}\gamma^{\mu}s),
\end{eqnarray}
where
\begin{eqnarray}
  \mathcal{C}_{D}\sim\frac{F_{D}(V_{1},V_{2})}{16},\quad
  \mathcal{C}_{B_d}\sim\frac{F_{Bd}(V_{1},V_{2})}{16},\quad
  \mathcal{C}_{B_s}\sim\frac{F_{Bs}(V_{1},V_{2})}{16},
\end{eqnarray}
and
\begin{eqnarray}
F_{D}(V_{1},V_{2})&=& F_{Bd}(V_{1},V_{2})=F_{K}(V_{1},V_{2}),\notag\\
F_{Bs}(V_{1},V_{2})&=&\frac{4}{3 (2 V_{1} +V_{2} )^2}+\frac{1}{6
(V_2-V_1) ^2}.
\end{eqnarray}
To the lowest order, we neglect the mixing matrices $U^u,U^d$,
and the same mixture of $Z_i$ in $F_2$ and $F_5$ lead to the
result $F_D=F_{Bs}=F_K$. Using data from~\cite{PDG_2012,fK_Rosner:2010ak,fD_Alexander:2009ux,Lellouch:2000tw,fB_Bernard:2009wr} we can get other lower bounds, which are list in the Table.\ref{tab:res}.
 \begin{table}
                \begin{tabular}{c | c  c  c  c }
    \hline
    $ P^0-\bar{P^0}$ & $[\Delta m_{meson}]^{PDG}$ 
    & $ M_{meson} $ & $f_{meson}$ &$ V_1 \ge$ \\
    \hline
     $ K-\bar{K}$ &$ (3.483\pm0.006)\times10^{-12}$
        &$497.6$ &$156\pm1.2$&$7.0\times10^7$ \\
     $ D-\bar{D}$     &$ (1.57^{+0.39}_{-0.41} )\times10^{-11}$
        &$1864.86\pm0.13$  &$206\pm11$&$8.4\times10^7$ \\
     $ B_d-\bar{B_d}$     &$ (3.337\pm0.033)\times10^{-10}$
        &$5279.58\pm0.17$  &$195\pm11$&$2.9\times10^7$ \\
     $ B_s-\bar{B_s}$     &$ (116.4  \pm0.5 )\times10^{-10}$
        &$5366.77\pm0.24$  &$243\pm11$&$0.7\times10^7$ \\
     \hline
        \end{tabular}
                \caption{Constrains on the family symmetry breaking scale $V_1$ from different neutral meson systems. The values are all in unit of $MeV$.}\label{tab:res}
        \end{table}

It's obvious from Table.\ref{tab:res} that the $K^0$-$\overline{K^0}$ system and
$D^0$-$\bar{D}^{0}$ system give the most stringent constraints on
$V_1$. The lower bounds turn out to be about $70\sim 84 $ TeV. $V_0$
can be got through $V_1$ with Eq.(\ref{eq:v0}), which turns out to be about $300$TeV. To apply seesaw mechanism at this scale, we need tuning the Yukawa
coupling to $10^{-4}$. Although not very nature, it's much better
than the situation in SM. It is notable that the constrains on the scales are not depend on the gauge coupling strength $g_F$. If we take it on the same order as the weak interaction, the mass of the new lightest gauge family boson can be about $100$TeV. This energy scale is at the reach of the next generation $100$TeV colliders.

\section{Semi-leptonic decay of Kaon}\label{sec:kaon_rare_decay} 

In SM FCNC processes occur at loop level through box diagrams and penguin diagrams~\cite{Buras:2009ye,Buras:2012fs}. These processes are suppressed by high order coupling, loop factor $1/16\pi^2$, and CKM factors in power of $\lambda\sim0.22$. With the new gauge bosons, FCNC process can happen at tree level.
 The new gauge bosons may manifest themselves and play a crucial roles in such processes. On the other hand, due to their heavy masses, there is almost no significant effect on the SM tree level
allowed channels. For example, the rare kaon decay process $K\rightarrow \pi \nu\bar{\nu }$, and LFNV
processes $\mu\rightarrow eee$.

In SM, the rare Kaon decay processes are induced by
Z-penguin diagram and box diagram. And the channel $K_L\rightarrow
\pi^0\nu\bar{\nu}$ violates CP directly~\cite{Grossman:1997sk}, providing same flavour
contents of the final neutrino pair.

The couplings between SM fermions and the new gauge bosons provide
several new $|\Delta S|=1$ low energy effective Hamiltonian terms, for
the final neutrinos with arbitrary flavour contents, the effective
Hamiltonian terms are:
\begin{eqnarray}
  \mathcal{H}_{eff}(K \rightarrow \pi\nu\bar{\nu})
  &=&\mathcal{C}_{lm}(\bar{s}\gamma_{\mu}d)\otimes
  (\bar{\nu}_l\gamma^{\mu}\nu_m)+h.c.
\end{eqnarray}
where $l,m={e,\mu,\tau}$, and the numerical values of the coefficient matrix elements
for  ${r_1}\sim\lambda$, ${r_2}\sim2\lambda$ are
\begin{eqnarray}
\mathcal{C}_{lm}= \left(
\begin{array}{ccc}
 -0.172 & -0.616 & -0.329 \\
 0.729 & 0.329 & 1.40 \\
 -0.248 & -1.54 & -0.157 \\
\end{array}
\right),
\end{eqnarray}
The diagonal matrix elements correspond to same flavour neutrino final
states. We can sum these channels incoherently and get the coefficient
being $\sum_l|\zeta_{ll}|^2\sim0.16$.

We only focus on the left-handed neutrinos, thus the leptonic current
takes a $V$-$A$ form. As for the hadronic current, since $\langle{\pi}|A^{\mu}|K\rangle=0$, the final result
only depends on $\langle{\pi}|V-A|K\rangle$. We have
\begin{eqnarray}
  \mathcal{H}_{eff}^{CP}&=&\frac{0.4}{8V_{0}^{2}}
  (\bar{s}d)_{V-A}(\bar{\nu_{\alpha}}\nu_{\alpha})_{V-A}+h.c.,
\end{eqnarray}
where the neutrino pairs belong to weak eigenstates and have the same flavour. Using the
isospin symmetry relation:
\begin{eqnarray}
&&\langle\pi^+|{(\bar{s}d)_{V-A}}|K^+\rangle=\sqrt{2}\langle\pi^0|{(\bar{s}u)_{V-A}}|K^+\rangle,\notag\\
&&\langle\pi^0|{(\bar{s}d)_{V-A}}|K_L\rangle=\sqrt{2}\langle\pi^0|{(\bar{s}u)_{V-A}}|K^+\rangle,
\end{eqnarray}
we have
\begin{eqnarray}
\frac{Br(K_L\rightarrow \pi^0\bar{\nu_{\alpha}}\nu_{\alpha})|_{New}}{Br(K^+\rightarrow \pi^0\nu_e e^+)|_{SM}}\sim
  \frac{Br(K^+\rightarrow \pi^+\bar{\nu_{\alpha}}\nu_{\alpha})|_{New}}{Br(K^+\rightarrow \pi^0\nu_e e^+)|_{SM}}
  \sim\left[ \frac{2\times0.4}{8 G_F V_{0}^{2} }
 \right]^2,
\end{eqnarray}
Taking $V_{0}\sim 3\times 10^2TeV$ and using the result
$Br(K^+\rightarrow \pi^0\nu e^+)=(5.07\pm0.04)\%$~\cite{PDG_2012},
we can get the  branch ratio
\begin{eqnarray}
  Br(K^+\rightarrow \pi^+\nu\bar{\nu})|_{New}\approx
  Br(K_L\rightarrow \pi^0\nu\bar{\nu})|_{New}
  \simeq
  4.6\times 10^{-16}.
\end{eqnarray}
The SM predicts these semi-leptonic FCNC processes have tiny
branch ratios~\cite{Buras:2001pn}
\begin{equation}\label{SMrare}
  \begin{split}
      &Br(K^+\rightarrow \pi^+\nu\bar{\nu})|_{SM}=(1.5^{+3.4}_{-1.2})\times10^{-10},\\
  &Br(K_L\rightarrow \pi^0\nu\bar{\nu})|_{SM}=(2.6\pm1.2)\times10^{-11}.
  \end{split}
\end{equation}
We find the contributions from new gauge bosons are far below the SM prediction in
Eq.(\ref{SMrare}). The CP violation in $K_L\rightarrow \pi
\bar{\nu}\nu$ is still dominated by SM contribution.


\section{Lepton flavour changing processes}
\label{sec:LFNV}

 In SM, LFNV processes are caused by the non-zero masses
of neutrinos~\cite{RevModPhys.MassiveNu} and neutrino mixing.
 There are several interesting LFNV processes, such as
$\mu^{-}\rightarrow e^{-}+\gamma$ and $\mu^{-}\rightarrow
e^{-}e^{+}e^{-}$. In SM, these processes are loop level effects
and highly suppressed. SM predictions of these processes are hopelessly small~\cite{LFNV},
\begin{eqnarray}
  \label{eq:mu2egamma}
 Br(\mu\rightarrow e \gamma)|_{SM}&\sim& 10^{-54},\notag\\
 Br(\mu\rightarrow e e e )|_{SM}&\sim&  10^{-56}.
\end{eqnarray}
The experimental bounds on the branch ratios at 90\% C.L. are~\cite{PDG_2012}
\begin{eqnarray}
&&Br(\mu^{-}\rightarrow
e^{-}\gamma)|_{Exp}<1.2\times 10^{-11}\\
&&Br(\mu^{-}\rightarrow
e^{-}e^{+}e^{-})|_{Exp}<1.0\times 10^{-12}.
\end{eqnarray}

The process $\mu\rightarrow e \gamma$ are not influenced by the new
gauge bosons at tree level. However, for $\mu^{-}\rightarrow e^{-}e^{+}e^{-}$, there are tree level
contributions mediated by the new gauge bosons. Here with the assumption that $U_{e}\sim
1$, we get the effective Hamiltonian for this process is
\begin{eqnarray}
  \mathcal{H}_{eff}(\mu\rightarrow 3e)=\frac{1}{8 V_{0}^{2}}\frac{F({r_1},{r_2})}{G({r_1},{r_2})}
  (\bar{e}\gamma_{\mu}\mu)\otimes
  (\bar{e}\gamma^{\mu}e)+H.c.,
\end{eqnarray}
where
\begin{eqnarray}
G({r_1},{r_2})=&216 {r_1} ^3-72 {r_1} ^2 {r_2} ^2+432 {r_1} ^2
{r_2} +198 {r_1} ^2-96
   {r_1}  {r_2} ^3+216 {r_1}  {r_2} ^2\notag\\
   &+264 {r_1}  {r_2}+60 {r_1} -24
   {r_2} ^4+16 {r_2} ^3+74 {r_2} ^2+40 {r_2} +6,\notag\\
 F({r_1},{r_2})=&-12 \sqrt{3} {r_1} ^3-24 {r_1} ^3+6 {r_1} ^2 {r_2} ^2-21 \sqrt{3} {r_1}
   ^2 {r_2} -51 {r_1} ^2 {r_2} -7 \sqrt{3} {r_1} ^2\notag\\
   &-14 {r_1} ^2+8 {r_1}
    {r_2} ^3-12 \sqrt{3} {r_1}  {r_2} ^2-32 {r_1}  {r_2} ^2-6 \sqrt{3}
   {r_1}  {r_2} -22 {r_1}  {r_2} \notag\\
   & -\sqrt{3} {r_1}-2 {r_1} +2 {r_2} ^4-3
   \sqrt{3} {r_2} ^3-5 {r_2} ^3-\sqrt{3} {r_2} ^2-8 {r_2} ^2-2 {r_2}.
\end{eqnarray}
Taking ${r_1}\sim\lambda$, ${r_2}\sim2\lambda$, we get
$F({r_1},{r_2})/G({r_1},{r_2})\sim -0.12$. The branching ratio for
this channel is
\begin{eqnarray}
Br(\mu\rightarrow 3e)=
\frac{\Gamma_{\mu\rightarrow3e}}{\Gamma_{\mu\rightarrow e \nu
\bar{\nu}}} \sim \left[ \frac{-0.12\sqrt{2}}{8 G_F V_{0}^{2} }
 \right]^2.
\end{eqnarray}
Assuming $V_{0}\geq 3\times10^2TeV$ , we get
\begin{eqnarray}
  Br(\mu\rightarrow 3e)\leq
  4.1\times10^{-16}.
\end{eqnarray}
This result is much larger than the SM prediction in Eq.(\ref{eq:mu2egamma}) but still
 below the experimental bound~\cite{PDG_2012}. The contribution of new physics in this process is of same order as that in $K_L \rightarrow \pi \bar{\nu}\nu$. Both of their
 initial flavours are changed. And they are induced by the mixing among the heavy family gauge bosons $F_{1},F_{4},F_{6}$
and $F_3,F_8$. There are many similar processes, such as the
rare B decays through $B\rightarrow X_s\mu^{-}\mu^{+}$, rare Kaon decay
through $K_L\rightarrow \pi^0 e^{+}e^{-}$, $K_L\rightarrow 
{\mu}^{+}{\mu}^{-}$.
Their branching ratios are of the same order, i.e. $10^{-16}$ from
the new gauge bosons' contributions. And they are all below the various experimental
bounds. These results make the lower bound $V_0 \sim 300TeV$ safe.

\section{conclusion}
\label{sec:conclusion}

We have investigated the structure of $SU(3)_F$
gauge family symmetry model and its low energy phenomenal results in flavour physics. This family
symmetry undergoes spontaneous breaking to $SO(3)_F$ and
then to a residual $Z_2$ symmetry. Seesaw mechanism is widely used
both in leptonic sector and quark sector to explain the observed
mass hierarchy and mixing structure, especially the neutrinos'
mass spectrum. The equality of seesaw scale and flavour symmetry breaking
scale needs a tuning of the Yukawa couplings, about $10^{-4}$,
which are much softer than SM. New scalar field is introduced and may be
a dark matter candidate. Also new CP violation phases appear and
may provide a solution to the baryon asymmetry in the universe.
The symmetry breaking mode makes the new gauge bosons can be divided
into two groups. Their mass scales can be constrained through the
mass differences of $P^0$-$\bar{P^0}$ meson systems. We get the
broken scale of the new gauge family symmetry is about
$V_{0}\geq 300$ TeV, and mass of the lightest new gauge boson can be
low as $100$ TeV. These new gauge bosons can induce FCNC processes
at tree level, and their contributions are suppressed by their heavy
masses and the resulting branching ratios are about
$10^{-16}$, which is $4\sim 5$ order below the current
experimental bounds. We expect the improvement of the rare FCNC
processes' measurements, as well as some exotic processes'
discovery, which may be found in the next running of LHC and the
next generation colliders of $100$ TeV, can throw some light upon
this new flavour symmetry.

\section*{Acknowledgments}
This work is supported in part by the National
Natural Science Foundation of China (NSFC) under Grants No. 10821504, 10975170, 11405095, and the Project of Knowledge Innovation Program (PKIP) of the Chinese Academy of Sciences. SSB would like to thank the China Scholarship Council.

\begin{appendix}
\section{} \label{sec:app} 

The field strengths of all gauge fields,
including the $SU(3)$ family symmetry, are defined as
\begin{eqnarray}
  F_{\mu\nu}^a&=&\partial_\mu A_{F,\nu}^a-\partial_\nu
  A_{F,\mu}^a+g_Ff^{abc}A_{F,\mu}^bA_{F,\nu}^c,\notag\\
    G_{\mu\nu}^a&=&\partial_\mu G_{\nu}^a-\partial_\nu
  G_{\mu}^a+g_s f^{abc}G_{\mu}^bG_{\nu}^c,\notag\\
   W_{\mu\nu}^a&=&\partial_\mu W_{\nu}^a-\partial_\nu
  W_{\mu}^a+g_w\epsilon^{abc}W_{\mu}^bW_{\nu}^c,\notag\\
   B_{\mu\nu}&=&\partial_\mu B_{\nu}-\partial_\nu
  B_{\mu}.
\end{eqnarray}
We define the covariant derivative as
\begin{eqnarray}
  D_\mu
&=&\partial_\mu-ig_FA_{F,\mu}^aT^a-ig_sG_\mu-g_{w}W_\mu+ig_{w}'YB_\mu\notag\\
&=&D_{\mu}^{SM}-ig_FA_{F,\mu}^aT^a.
\end{eqnarray}
The full Lagrangian is
\begin{eqnarray}
  \mathcal{L}=\mathcal{L}_G+\mathcal{L}_{k}+\mathcal{L}_{H}+\mathcal{L}_{Y}+\mathcal{L}_{N},
\end{eqnarray}
with each term defined as follows
\begin{eqnarray}
  \label{eq:LG}
  \mathcal{L}_{G}&=&-\frac{1}{4}\left(F_{\mu\nu}^aF^{a\mu\nu}+G_{\mu\nu}^bG^{b\mu\nu}+W_{\mu\nu}^cW^{c\mu\nu}
+B_{\mu\nu}B^{\mu\nu}\right)\\
  \mathcal{L}_{k}&=&\overline{u}_{L,R}i\gamma^\mu
  D_\mu u_{L,R}+\overline{d}_{L,R}i\gamma^\mu
  D_\mu d_{L,R}+\overline{e}_{L,R}i\gamma^\mu
  D_\mu e_{L,R}+\overline{\nu}_{L}i\gamma^\mu
  D_\mu\nu_{L},\notag\\
   \label{eq:Lint}\\
  \mathcal{L}_{H}&=&\mathcal{L}_{DH}-V[H,\Phi_1,\Phi_2,\Phi_\nu,\phi_s]\notag\\
  &=&\left(D_{\mu}^{SM}H\right)^\dagger \left(D^{\mu,SM}H\right) + \tr \left(D_\mu\Phi_1
  (D^\mu\Phi_1)^\dagger\right)+\tr \left(D_\mu\Phi_2
  (D^\mu\Phi_2)^\dagger\right)\notag\\
&&+
  \tr \left(D_\mu\Phi_\nu (D^\mu\Phi_\nu)^*\right)+\partial_\mu\phi_s\partial^\mu\phi_s
-V\left(H,\Phi_1,\Phi_2,\Phi_\nu,\phi_s\right). \label{eq:Lhiggs}\\
\mathcal{L}_{Y}&=&y_L^{u}\bar{l}H
U+y_R^u\overline{u}_{R}\phi_s
 U+\frac{1}{2}\overline{U}(\Delta_{1}^{U}\Phi_1+\Delta_{2}^{U}\Phi_2) U\notag\\
&&+ y_L^{d}\bar{l}\tilde{H} D+y_R^d\overline{d}_{R}\phi_s
D+\frac{1}{2}\overline{D}(\Delta_{1}^{D}\Phi_1+\Delta_{2}^{D}\Phi_2) D\notag\\
&&+ y_L^{e}\bar{l}H E+y_R^e\overline{e}_R\phi_s
E+\frac{1}{2}\overline{E}(\Delta_{1}^{E}\Phi_1+\Delta_{2}^{E}\Phi_2) E\notag\\
&&+y_L^{\nu}\bar{l}\tilde{H}N_R+\frac{1}{2}\xi^\nu\overline{N}_R\Phi_\nu
N_{R}^{c}+H.C..\\
  \mathcal{L}_{N}&=&
  i\overline{U}\gamma^\mu(\partial_\mu-ig_s G_{\mu}-ig_FA_{F,\mu}^aT^a)U
  +i\overline{D}\gamma^\mu(\partial_\mu-ig_s G_{\mu}-ig_FA_{F,\mu}^aT^a)D\notag\\
  &&+i\overline{E}\gamma^\mu(\partial_\mu-ig_FA_{F,\mu}^aT^a)E
  +i\overline{N_R}\gamma^\mu(\partial_\mu-ig_FA_{F,\mu}^aT^a)N_R.
  \label{eq:Lnew}
\end{eqnarray}
\end{appendix}

\end{document}